\documentclass{ws-procs975x65}
\usepackage{graphicx}
\def\rightnote{}
\begin{document}

\title{The Thermodynamic Arrow: Puzzles and Pseudo-puzzles}

\author{Huw Price}

\address{Centre for Time, Main Quad A14,
University of Sydney, NSW 2006, Australia.\\ Department of Philosophy, University of Edinburgh, David Hume Tower,\\
George Square, Edinburgh EH8 9JX,
Scotland.\\E-mail:
huw@mail.usyd.edu.au}

\maketitle

\abstracts{
For more than a century, physics has known of a puzzling 
conflict between the T-asymmetry of thermodynamic phenomena 
and the T-symmetry of the underlying microphysics on which 
these phenomena depend. This paper provides a guide to the current 
status of this puzzle, distinguishing the central issue from 
various issues with which it may be confused. It is shown that 
there are two competing conceptions of what is needed to resolve 
the puzzle of the thermodynamic asymmetry, which differ with 
respect to the number of distinct T-asymmetries they take 
to be manifest in the physical world. On the preferable one-asymmetry 
conception, the remaining puzzle concerns the ordered distribution 
of matter in the early universe. The puzzle of the thermodynamic 
arrow thus becomes a puzzle for cosmology.}

\section{The puzzle of temporal bias}
Late in the nineteenth century, on the shoulders of Maxwell, 
Boltzmann and many lesser giants, physicists saw that 
there is a deep puzzle behind the familiar phenomena described 
by the new science of thermodynamics. On the one hand, many such
 phenomena show a striking temporal bias. They are common 
in one temporal orientation, but rare or non-existent in reverse. 
On the other hand, the underlying laws of mechanics show no 
such temporal preference. If they allow a process in one direction, 
they also allow its temporal mirror image. Hence the puzzle: 
if the laws are so even-handed, why are the phenomema themselves 
so one-sided?

What has happened to this puzzle since the 1890s? I suspect that 
many contemporary physicists regard it as a dead issue, long 
since laid to rest. Didn't it turn out to be just a matter of 
statistics, after all? However, while there are certainly would-be 
solutions on offer---if anything, as we'll see, too many of them---it 
is far from clear that the puzzle has actually been solved. Late 
in the twentieth century, in fact, one of the most authoritative 
writers on the conceptual foundations of statistical mechanics 
could still refer to an understanding of the time-asymmetry of 
thermodynamics as `that obscure object of desire'.\cite{sklar}

One of the obstacles to declaring the problem solved is that 
there are several distinct approaches, not obviously compatible 
with one another. Which of these, if any, is supposed to be \textit{the} 
solution, now in our grasp? Even more interestingly, it turns 
out that not all these would-be solutions are answers to the 
same question. There are different and incompatible conceptions 
in the literature of  what the puzzle of the thermodynamic asymmetry 
actually is---about what \textit{exactly} we should be trying explain, 
when we try to explain the thermodynamic arrow of time.

What the problem needs is therefore what philosophers do for 
a living: drawing fine distinctions, sorting out ambiguities, 
and clarifying the logical structure of difficult and subtle 
issues. My aim here is to bring these methods to bear on the 
puzzle of the time-asymmetry of thermodynamics. I want to distinguish 
the true puzzle from some of the appealing false trails, and 
hence to make it clear where physics stands in its attempt to 
solve it. 

Little here is new, but it is surprisingly 
difficult to find a clear guide to these matters in the literature, 
either in philosophy or in physics. Accordingly, I think the 
paper will serve a useful purpose, in helping non-specialists 
to understand the true character of the puzzle discovered by 
those nineteenth century giants, the extent to which it has been 
solved, and the nature of the remaining issues.\footnote{For those
interested in more details, I discuss these topics at greater length
elsewhere.\cite{hp1,hp2,hp3}}

\section{The true puzzle---a first approximation and a popular challenge}

Everyone agrees, I think, that the puzzle of the thermodynamic 
arrow stems from the conjunction of two facts (or apparent facts---one 
way to dissolve the puzzle would be to show that one or other 
of the following claims isn't actually true):

\begin{enumerate}
\item There are many common and familiar physical processes, collectively 
describable as cases in which entropy is increasing, whose corresponding 
time-reversed processes are unknown or at least very rare.
\item The dynamical laws governing such processes show no such T-asymmetry---if 
they permit a process to occur with one temporal orientation, 
they permit it to occur with the reverse orientation.
\end{enumerate}

\noindent As noted, some people will be inclined to object at this point
that the conjunction is  \textit{merely}  apparent. In particular, it may
be objected that we now know that the dynamical laws are not
time-symmetric. Famously, T-symmetry is violated in weak interactions, by
the neutral
$K$ meson. Doesn't this eliminate the puzzle?

No. If the time-asymmetry of thermodynamics were associated with 
the T-symmetry violation displayed by the neutral \textit{K} meson, then
anti-matter  would show the reverse of the normal thermodynamic
asymmetry.  Why? Because PCT-symmetry guarantees that if we replace
matter  by anti-matter (i.e., reverse P and C) and then view the result 
in reverse time (i.e., reverse T), physics remains the same. 
So if we replaced matter by anti-matter but didn't reverse time, 
any intrinsic temporal arrow or T-symmetry violation would reverse its 
apparent direction. In other words, physicists in anti-matter 
galaxies find the opposite violations of T-symmetry in weak interactions 
to those found in our galaxy. So if the thermodynamic arrow were 
tied to the T-symmetry violation, it too would have to reverse under such 
a transformation.

But now we have both an apparent falsehood, and a paradox. There's 
an apparent falsehood because (of course) we don't think that 
anti-matter behaves anti-thermodynamically. We expect stars in 
anti-matter galaxies to radiate just like our own sun (as the 
very idea of an anti-matter galaxy requires, in fact). And there's 
a paradox, because if this were the right story, what would happen 
to particles which are their own anti-particles, such as photons? 
They would have to behave both thermodynamically and anti-thermodynamically!

Here's another way to put the point. The thermodynamic arrow 
isn't just a T-asymmetry, it is a PCT-asymmetry as well. There 
are many familiar process whose PCT-reversed processes are equally 
compatible with the underlying laws, but which never happen, 
in our experience. We might be tempted to explain this asymmetry 
as due to the imbalance between matter and anti-matter, but the 
above reflections show that this is not so. So instead of the 
puzzle of the T-asymmetry of thermodynamics, we could speak of 
the puzzle of the PCT-asymmetry of thermodynamics. Then it would 
be clear to all that the strange behaviour of the neutral \textit{K} 
meson isn't relevant. Knowing that we could if necessary rephrase 
the problem in this way, we can safely rely on the simpler formulation, 
and return to our original version of the puzzle.

\section{Four things the puzzle is not}

Some of the confusions common in debates about the origins of 
the thermodynamic asymmetry can be avoided distinguishing the 
genuine puzzle from various pseudo-puzzles with which it is liable 
to be confused. In this section I'll draw four distinctions of 
this kind.

\subsection{The meaning of irreversibility}

The thermodynamic arrow is often described in terms of the `irreversibility' 
of many common processes---e.g., of 
what happens when a gas disperses from a pressurised bottle. This makes it sound as if the problem is that we can't 
make the gas behave in the opposite way---we can't make it put 
itself back into the bottle. Famously, Loschmidt's reversibility 
objection rested on pointing out that the reverse motion 
is equally compatible with the laws of mechanics. 
Some responses to this problem concentrate on the issue as to 
why we can't actually reverse the motions (at least in most
cases).\cite{ridd}

This response misses the interesting point, however. The interesting 
issue turns on a numerical imbalance in nature between `forward' 
and `reverse' processes, not case-by-case irreversibility of individual 
processes. Consider a parity analogy. Imagine a world containing 
many left hands but few right hands. Such a world shows an interesting 
parity asymmetry, even if any individual left hand can easily 
be transformed into a right hand. Conversely, a world with equal 
numbers of left and right hands is not interestingly P-asymmetric, 
even if any individual left or right hand cannot be reversed. 
Thus the interesting issue concerns the numerical asymmetry between 
the two kinds of structures---here, left hands and right hands---not 
the question whether one can be transformed into the other.

Similarly in the thermodynamic case, in my view. The important 
thing to explain is the numerical imbalance in nature between 
entropy-increasing processes and their T-reversed counterparts, 
not the practical irreversibility of individual processes.

\subsection{Asymmetry in time versus asymmetry of time}

Writers on the thermodynamic asymmetry often write as if the 
problem of explaining this asymmetry is the problem of explaining 
`the direction of time'. This may be a harmless way of speaking, 
but we should keep in mind that the real puzzle concerns the 
asymmetry of physical processes \textit{in} time, not an 
asymmetry \textit{of time itself.} By analogy, imagine a long narrow 
room, architecturally symmetrical end-to-end. Now suppose all 
the chairs in the room are facing the same end. Then there's 
a puzzle about the asymmetry in the arrangement of the chairs, 
but not a puzzle about the asymmetry of the room. Similarly, 
the thermodynamic asymmetry is an asymmetry of the `contents' 
of time, not an asymmetry of the container itself.

It may be helpful to make a few remarks about the phrase `direction 
of time'. Although this expression is in common use, it isn't 
at all clear what it could actually mean, if we try to take it 
literally. Often the thought seems to be that there is an objective 
sense in which one time direction is future (or `positive'), and 
the other past (or `negative'). But what could this distinction 
amount to? It's easy enough to make sense of idea that time is \textit{anisotropic}---i.e., 
different in one direction than in the other. For example, time 
might be finite in one direction but infinite in the other. But 
this isn't enough to give a \textit{direction} to time, in above sense. 
After all, if one direction were objectively the future or positive 
direction, then in the case of a universe finite at one end, 
there would be two possibilities. Time might be finite in the 
past, and or finite in the future. So anisotropy alone doesn't 
give us \textit{direction}.

Similarly, it seems, for any other physical time-asymmetry 
to which we might appeal. If time did have a direction---an objective 
basis for a privileged notion of positive or future time---then 
for any physical arrow or asymmetry in time, there would always 
be a question as to whether that arrow pointed forwards or 
backwards. And so no physical fact could answer this question, 
because for any candidate, the same issue arises all over again. 
Thus the idea that time has a real direction seems without any 
physical meaning. (Of course, we can use any asymmetry we like 
as a basis for a conventional labelling---saying, for example, that 
we'll regard the direction in which entropy is increasing as 
the positive direction of time. But this is different from discovering 
some intrinsic directionality to time itself.)

For present purposes, then, I'll assume that it is a conventional 
matter which direction we treat as positive or future time. Moreover, 
although it makes sense to ask whether time is anisotropic, it 
seems clear that this is a different issue from that of the thermodynamic 
asymmetry. As noted, the thermodynamic asymmetry is an asymmetry 
of physical processes \textit{in} time, not an asymmetry of time itself.

\subsection{Entropy gradient not entropy increase}

If it is conventional which direction counts as positive 
time, then it is also conventional whether entropy increases 
or decreases. It increases by the lights of the usual convention, 
but decreases if we reverse the labelling. But this may seem 
ridiculous. Doesn't it imply, absurdly, that the thermodynamic 
asymmetry is merely conventional?

No. The crucial point is that while it's a conventional matter 
whether the entropy gradient slopes up or down, the gradient 
itself is objective. The puzzling asymmetry is that the gradient 
is monotonic---it slopes in the same direction everywhere (so 
far as we know).

It is worth noting that in principle there are two possible ways 
of contrasting this monotonic gradient with a symmetric world. 
One contrast would be with a world in which there are entropy 
gradients, but sometimes in one direction and sometimes in the 
other---i.e., worlds in which entropy sometimes goes up and sometimes 
goes down. The other contrast would be with worlds in which there 
are no significant gradients, because entropy is always high. 
If we manage to explain the asymmetric gradient we find in our 
world, we'll be explaining why the world isn't symmetric in one 
of these ways---but which one? The answer isn't obvious in advance, 
but hopefully will fall out of a deeper understanding of the 
nature of the problem.

\subsection{The term `entropy' is inessential }
A lot of time and ink has been devoted to the question how entropy 
should be defined, or whether it can be defined at all in certain 
cases (e.g., for the universe as a whole). It would be easy to 
get the impression that the puzzle of the thermodynamic asymmetry 
depends on all this discussion---that whether there's really a 
puzzle depends on how, and whether, entropy can be defined, perhaps.

But in one important sense, these issues are beside the point. 
We can see that there's a puzzle, and go a long way towards saying 
what it is, without ever mentioning entropy. We simply need to 
describe in other terms some of the many processes which show 
the asymmetry---which occur with one temporal orientation but 
not the other. For example, we can point out that there are lots 
of cases of big difference in temperatures spontaneously equalising, 
but none of big differences in temperature spontaneously arising. 
Or we can point out that there are lots of cases of pressurised 
gas spontaneously leaving a bottle, but none of gas spontaneously 
pressurising by entering a bottle. And so on.

In the end, we may need the notion of entropy to generalise properly 
over these cases. However, we don't need it to see that there's 
a puzzle---to see that there's a striking imbalance in nature between 
systems with one orientation and systems with the reverse orientation. 
For present purposes, then, we can ignore objections based on 
problems in defining entropy. (Having said that, of course, we 
can go on using the term entropy with a clear conscience, without 
worrying about how it's defined. In what follows, talk of entropy increase is just 
a placeholder for a list of the actual phenomena which display 
the asymmetry we're interested in.)

\subsection{Summary}
For the remainder of the paper, then, I take it (i) that the 
asymmetry in nature is a matter of numerical imbalance between 
temporal mirror images, not of literal reversibility; (ii) that 
we are concerned with an asymmetry of physical processes in time, 
not with an asymmetry in time itself; (iii) that the objective 
asymmetry concerned is a monotonic \textit{gradient,} rather 
than an increase or a decrease; and (iv) that if need be the 
term `entropy' is to be thought of as a placeholder for 
the relevant properties of a list of actual physical asymmetries.

\section{What would a solution look like? Two models}
With our target more clearly in view, I now want to call attention 
to what may be the most useful distinction of all, in making 
sense of the many things that physicists and philosophers say 
about the thermodynamic asymmetry. This is a distinction between 
two very different conceptions of \textit{what it would take} to explain 
the asymmetry---so different, in fact, that they disagree on \textit{how 
many} distinct violations of T-symmetry it takes to explain the 
observed asymmetry. On one conception, an explanation needs two 
T-asymmetries. On the other conception, it needs only one.

Despite this deep difference of opinion about what 
a solution would look like, the distinction between these two 
approaches is hardly ever noted in the literature---even by philosophers, 
who are supposed to have a nose for these things. So it is easy 
for advocates of the different approaches to fail to see that 
they are talking at cross-purposes---that in one important sense, 
they disagree about what the problem is.

\subsection{The two-asymmetry approach}

Many approaches to the thermodynamic asymmetry look for a dynamical 
explanation of the second law---a dynamical cause or factor, responsible 
for entropy increase. Here are some examples, old and new:

\begin{enumerate}
\item \textit{The H-theorem.} Oldest and most famous of all, 
this is Boltzmann's development of Maxwell's idea that intermolecular 
collisions drive gases towards equilibrium.
\item \textit{Interventionism.} This alternative to the \textit{H}-theorem, apparently 
first proposed by S. H. Burbury in the
1890s,\cite{burbury1,burbury2} attributes entropy  increase to the
effects of random and uncontrollable influences  from a system's external
environment.
\item \textit{Indeterministic dynamics.} There are various attempts to 
show how an indeterministic dynamics might account for the second 
law. A recent example is a proposal that the stochastic collapse 
mechanism of the GRW approach to quantum theory might
also explain entropy increase.\cite{al1,al2}

\end{enumerate}

I stress two points about these approaches. First, if there is 
something dynamical which makes entropy increase, 
then it needs to be time-asymmetric. Why? Because otherwise it 
would force entropy to increase (or at least not to decrease) 
in both directions---in other words, entropy would be constant. 
In the \textit{H}-theorem, for example, this asymmetry resides in the 
assumption of molecular chaos. In interventionism, it is provided 
by the assumption that incoming influences from the environment 
are `random', or uncorrelated with the system's internal dynamical 
variables.

The second point to be stressed is that this asymmetry alone 
isn't sufficient to produce the observed thermodynamic phenomena. 
Something which forces entropy to be non-decreasing won't produce 
an entropy gradient unless entropy starts low. To give us the 
observed gradient, in other words, this approach also needs a 
low entropy boundary condition---entropy has to be low in the 
past. This condition, too, is time-asymmetric, and it's a separate 
condition from the dynamical asymmetry. (It is not guaranteed 
by the assumption of molecular chaos, for example.) 

So this approach is committed to the claim that it takes \textit{two} 
T-asymmetries---one in the dynamics, and one in the boundary conditions---to 
explain the observed asymmetry of thermodynamic phenomena. If 
this model is correct, explanation of the observed asymmetry 
needs an explanation of both contributing asymmetries, and the 
puzzle of the thermodynamic arrow has become a double puzzle.

\subsection{The one-asymmetry model}
The two-asymmetry model isn't the only model on offer, however. 
The main alternative was first proposed by Boltzmann in the
1870s,\cite{b02}  in response to Loschmidt's famous criticism of the
\textit{H}-theorem.  To illustrate the new approach, think of a large
collection of  gas molecules, isolated in a box with elastic walls. If
the motion  of the molecules is governed by deterministic laws, such as
Newtonian  mechanics, a specification of the microstate of the system at 
any one time uniquely determines its entire trajectory. The key 
idea of Boltzmann's new approach is that in the overwhelming 
majority of possible trajectories, the system spends the 
overwhelming majority of the time in a high entropy macrostate---among 
other things, a state in which the gas is dispersed throughout 
the container. (Part of Boltzmann's achievement was to find the 
appropriate way of counting possibilities, which we can call 
the \textit{Boltzmann measure.})

Importantly, there is no temporal bias in this set of possible 
trajectories. Each possible trajectory is matched by its time-reversed 
twin, just as Loschmidt had pointed out, and the Boltzmann measure 
respects this symmetry. Asymmetry arises only when we apply 
a low entropy condition at one end. For example, suppose we stipulate 
that the gas is confined to some small region at the initial 
time $t_{0}$. Restricted to the remaining trajectories, the Boltzmann 
measure now provides a measure of the likelihood of the various 
possibilities consistent with this boundary condition. Almost 
all trajectories in this remaining set will be such that the 
gas disperses after $t_{0}$. The observed behaviour is thus predicted 
by the time-symmetric measure, once we conditionalise on the 
low entropy condition at $t_{0}$.

On this view, then, there's no time-asymmetric factor which causes 
entropy to increase. This is simply the most likely thing to 
happen, given the combination of the time-symmetric Boltzmann 
probabilities and the single low entropy restriction in the past. 
More below on the nature and origins of this low entropy 
boundary condition. For the moment, the important thing is that 
although it is is time-asymmetric, so far as we know, this is 
the only time-asymmetry in play, according to Boltzmann's statistical 
approach. There's no need for a second asymmetry in the dynamics.

\section{Which is the right model?}
It is important to distinguish these two models, but it would 
be even more useful to know which of them is right. How many 
time-asymmetries should we be looking for, in trying to account 
for the thermodynamic asymmetry? This is a big topic, but I'll 
mention two factors, both of which seem to me to count in favour 
of the one-asymmetry model.

The first factor is simplicity, or theoretical economy. If the 
one-asymmetry approach works, it simply does more with less. 
In particular, it leaves us with only one time-asymmetry to explain. 
True, this would not be persuasive if the two-asymmetry approach 
actually achieved more than the one-asymmetry approach---if the 
former had some big thoeretical advantage that the latter lacked. 
But the second argument I want to mention suggests that this 
can't be the case. On the contrary, the second asymmetry seems 
redundant.

Redundancy is a strong charge, but consider the facts. The two-asymmetry approach tries to identify some dynamical factor (collisions, 
or external influences, or whatever) that causes entropy to increase---that 
makes a pressurised gas leave a bottle, for example. However, 
to claim that one of these factors \textit{causes} the gas to disperse 
is to make the following `counterfactual' claim: \textit{If the 
factor were absent, the gas would not disperse} (or would do so 
at a different rate, perhaps). But how could the absence of collisions 
or external influences \textit{prevent} the gas molecules from leaving 
the bottle?

Here's a way to make this more precise. In the terminology of 
Boltzmann's statistical approach, we can distinguish between \textit{normal} 
initial microstates (for a system, or for the universe as a whole), 
which lead to entropy increases much as we observe, and \textit{abnormal} 
microstates, which are such that something else happens. The 
statistical approach rests on the fact that normal microstates 
are vastly more likely than abnormal microstates, according to 
the Boltzmann measure.

In these terms, the above point goes as follows. The two-asymmetry 
approach is committed to the claim that the universe begins in 
an abnormal microstate. Why? Because in the case of normal initial 
microstates, entropy increases anyway, without the mechanism 
in question---so the required counterfactual claim isn't true.

It is hard to see what could justify this claim about the initial 
microstate. At a more local level, why should we think that the 
initial microstate of a gas sample in an open bottle is normally such 
that if it weren't for collisions (or external influences, or 
whatever), the molecules simply wouldn't encounter the open top 
of the bottle, and hence disperse?

Thus it is doubtful whether there is really any need for a dynamical 
asymmetry, and the one-asymmetry model seems to offer the better 
conception of what it would take to solve the puzzle of the thermodynamic 
asymmetry. But if so, then the various two-asymmetry approaches---including 
Boltzmann's own \textit{H}-theorem, which he himself defended in the 1890s, 
long after he first proposed the statistical approach---are looking 
for a solution to the puzzle in the wrong place, at least in 
part.

For present purposes, the main conclusion I want to emphasise 
is that we need to make a choice. The one-asymmetry model and 
the two-asymmetry model represent are two very different views 
of \textit{what it would take} to explain the thermodynamic arrow---of 
what the problem is, in effect. Unless we notice that they are 
different approaches, and proceed to agree on which of them we 
ought to adopt, we can't possibly agree on whether the old puzzle 
has been laid to rest.

\section{The Boltzmann-Schuetz hypothesis---a no-asymmetry solution? }
If the one-asymmetry view is correct,  the puzzle of the thermodynamic 
arrow is really the puzzle of the low entropy boundary condition. 
Why is entropy so low in the past? After all, in making it unmysterious 
why entropy doesn't decrease in one direction, the Boltzmann 
measure equally makes it mysterious why it does decrease in the 
other---for the statistics themselves are time-symmetric.

Boltzmann himself was one of the first to see 
the importance of this issue. In a letter to \textit{Nature} 
in 1895, he suggests an explanation, based on an idea he
attributes  to `my old assistant, Dr Schuetz'.\cite{b01} He notes that although
low  entropy states are very unlikely, they are very likely to occur 
eventually, given enough time. If the universe is very old, it 
will have had time to produce the kind of low entropy region 
we find ourselves inhabiting simply by accident. `Assuming the 
universe great enough, the probability that such a small part 
of it as our world should be in its present state, is no longer 
small,' as Boltzmann puts it.

\begin{figure}[htbp]
\begin{center}
\includegraphics[width=3.5in]{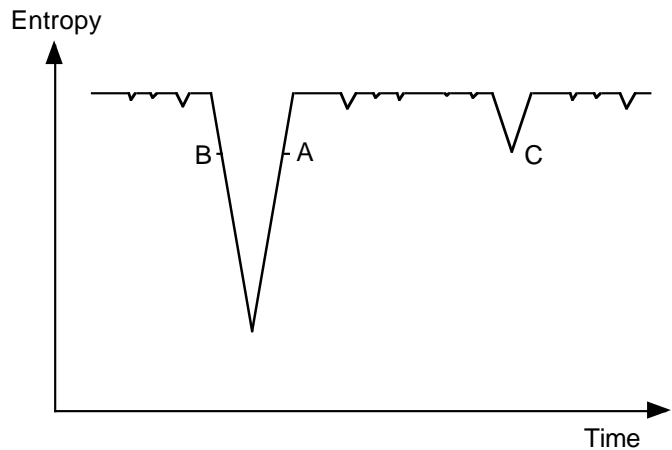}
\caption{Boltzmann's entropy curve.}
\label{default}
\end{center}
\end{figure}

It is one thing to explain why the universe contains regions 
like ours, another to explain why we find ourselves in such a 
region. If they are so rare, isn't it more likely that we'd find 
ourselves somewhere else? But Boltzmann suggests an answer to 
this, too. Suppose, as seems plausible, that creatures like us 
couldn't exist in the vast regions of near-equilibrium between 
such regions of low entropy. Then it's no surprise that we find 
ourselves in such an unlikely place. As Boltzmann himself puts 
it, `the ... \textit{H} curve would form a representation of what takes 
place in the universe. The summits of the curve would represent 
the worlds where visible motion and life exist.'

Figure 1 shows what Boltzmann calls the \textit{H} curve, except
that  this diagram plots entropy rather than Boltzmann's quantity
\textit{H.}  Entropy is low when \textit{H} is high, so the summits of
Boltzmann's \textit{H}  curve are the troughs of the entropy curve. The
universe spends  most of its time very close to equilibrium.
But  occasionally---much more rarely than this diagram actually
suggests---a  random re-arrangement of matter produces a state of low
entropy.  As the resulting state returns to equilibrium, there's an
entropy  slope, such as the one on which we (apparently) find ourselves, 
at a point such as A.

Why do we find ourselves on an uphill rather than a downhill 
slope, as at B? In another paper, Boltzmann offers a remarkable 
proposal to explain this, too.\cite{b03} Perhaps our perception of past 
and future depends on the entropy gradient, in such a way that 
we are bound to regard the future as lying `uphill'. Thus the 
perceived direction of time would not be objective, 
but a product of our own orientation in time. 
Creatures at point B would see the future as lying in the other 
direction, and there's no objective sense in which they are wrong 
and we are right, or vice versa. Boltzmann compares this to the 
discovery that spatial up and down are not absolute directions, 
the same for all observers everywhere.

For present purposes, what matters about the Boltzmann-Schuetz hypothesis 
is that it offers an explanation of the local asymmetry of thermodynamics 
in terms which are symmetric on a larger scale. So it is a no-asymmetry 
solution---the puzzle of the thermodynamic asymmetry simply vanishes 
on the large scale.

\section{The big problem }

Unfortunately, however, this clever proposal has a sting in its 
tail, a sting so serious that it now seems almost impossible 
to take the hypothesis seriously. The problem flows directly 
from Boltzmann's own link between entropy and probability. In 
Figure 1, the vertical axis is a logarithmic probability scale. 
For every downward increment, dips in the curve of the corresponding 
depth are exponentially more improbable. So a dip of the depth 
of point A or point B is much more likely to occur in the form 
shown at point C---where the given depth is very close to the 
minimum of the fluctuation---than in association with a much bigger 
dip, as at A and B. Hence if our own region has a past of even 
lower entropy, it is much more improbable than it needs to 
be, given its present entropy. So far, this point seems to have 
been appreciated already in the 1890s, in exchanges between Boltzmann 
and Zermelo. What doesn't seem to have appreciated is its devastating 
consequence, namely, that according to the Boltzmann measure 
it is much easier to produce fake records and memories, than 
to produce the real events of which they purport to be records.

Why does this consequence follow? Well, imagine that the universe 
is vast enough to contain many separate fluctuations, each containing 
everything that we see around us, including the complete works 
of Shakespeare, in all their twenty-first century editions. Now 
imagine choosing one of these fluctuations at random. It is vastly 
more likely that we'll select a case in which the Shakespearean 
texts are a product of a spontaneous recent fluctuation, than 
one in which they were really written four hundred years earlier 
by a poet called William Shakespeare. Why? Simply because entropy 
is much higher now than it was in the sixteenth century (as we 
normally assume that century to have been). Recall that according 
to Boltzmann, probability increases exponentially with entropy. 
Fluctuations like our twenty-first century---`Shakespearian' texts 
and all---thus occur much more often in typical world-histories 
than fluctuations like the lower-entropy sixteenth century. So 
almost all fluctuations including the former don't include the 
latter. The same goes for the rest of history---all our `records' 
and `memories' are almost certainly misleading.

To make this conclusion vivid we can take advantage of the fact 
that in the Boltzmann picture, there isn't an objective direction 
of time. So we can equally well think about the question of `what 
it takes' to produce what we see around us from the reverse of 
the normal temporal perspective. Think of starting in what we 
call the future, and moving in the direction we call towards 
the past. Think of all the apparently miraculous accidents it 
takes to produce the kind of world we see around us. Among other 
things, our bodies themselves, and our editions of Shakespeare, 
have to `undecompose', at random, from (what we normally think 
of as) their future decay products. That's obviously extremely 
unlikely, but the fact that we're here shows that it 
happens. But now think of what it takes to get even further 
back, to a sixteenth century containing Shakespeare himself. 
The same kind of near-miracle needs to happen many more times. 
Among other things, there are several billion intervening humans 
to `undecompose' spontaneously from dust.

So the Boltzmann-Schuetz hypothesis implies that our apparent 
historical evidence is almost certainly unreliable. So far as 
I know, this point is first made in print by von Weizs\"{a}cker 
in 1939.\cite{vw} Von Weizs\"{a}cker notes that `improbable states can
count  as documents [i.e., records of the past] only if we presuppose 
that still less probable states preceded them.' He concludes 
that `the most probable situation by far would be that the present 
moment represents the entropy minimum, while the past, which 
we infer from the available documents, is an illusion.' 

Von Weizs\"{a}cker also notes that there's another problem of a 
similar kind. The Boltzmann-Schuetz hypothesis implies that as 
we look further out into space, we should expect to find no more 
order than we already have reason to believe in. But we can now 
observe vastly more of the universe than was possible in Boltzmann's 
day, and there seems to be low entropy all the way out.

So the Boltzmann-Schuetz hypothesis faces some profound objections. 
Fortunately, as we're about to see, modern cosmology goes at 
least some way to providing us with an alternative.

\section{Initial smoothness}
We have seen that the observed thermodynamic asymmetry requires 
that entropy was low in the past. Low entropy requires concentrations 
of energy in useable forms, and presumably there are many ways 
such concentrations could exist in the universe. On the face 
of it, we seem to have no reason to expect any particularly neat 
or simple story about how it works in the real world---about where 
the particular concentrations of energy we depend on happen to originate.
Remarkably,  however, modern cosmology suggests that all the observed low 
entropy is associated with a single characteristic of the early universe, 
soon after the big bang. The crucial thing is that matter is distributed extremely smoothly 
in the early universe. This provides a vast 
reservoir of low entropy, on which everything else depends. In 
particular, smoothness is necessary for galaxy and star formation, 
and most familiar irreversible phenomena depend on the sun.

Why does a smooth arrangement of matter amount to a low entropy state?
Because in a system dominated by an attractive  force such as gravity, a
uniform distribution of matter is highly unstable (and provides a highly
useable supply of potential energy). However, about
$10^{5}$ years after the big bang, matter seems  to have been distributed
smoothly to very high accuracy. 

One way to get  a sense how surprising this is, is to
recall that we've found no reason  to disagree with Boltzmann's
suggestion that there's no objective  distinction between past and
future---no sense in which things  really happen in the direction we
think of as past-to-future.  Without such a distinction, there's no
objective sense in which  the big bang is not equally the end point of a
gravitational  collapse. Somehow that collapse is coordinated with
astounding  accuracy, so that the matter involved manages to avoid 
forming large agglomerations (in fact, black holes), and instead 
spreads itself out very evenly across the universe. (By calculating the entropy of black holes with comparable mass, 
Penrose\cite{penrose2} has  estimated the probability of such a smooth
arrangement of matter  at $10^{-10^{123}}$.)

In my view, this discovery about the cosmological origins of low entropy 
is one of the great  achievements of late twentieth century physics. 
It is a remarkable discovery in two quite distinct ways, in fact. First, it is the only anomaly
necessary  to account for the low entropy we find in the universe, at
least  so far as we know. So it is a remarkable theoretical achievement---it wraps up the entire puzzle of the
thermodynamic asymmetry into a single package, in effect. Second, it is astounding that it  happens at all, according to existing theories of how
gravitating  matter should behave (which suggests, surely, that there is something very important missing from those theories).\footnote{True, it is easy to fail to see how astounding the smooth early universe is, by failing to see that the big bang can quite properly be regarded as the end point of a gravitational collapse. But anyone inclined to deny the validity of this way of viewing the big bang faces a perhaps even more daunting challenge: to explain what is meant by, and what is the evidence for, the claim that time has an objective direction!}

\section{Open questions}

Why is the universe smooth soon after the big bang? This is a 
major puzzle, but---if we accept that the one-asymmetry model---it is 
the only question we need to answer, to solve 
the puzzle of the thermodynamic arrow. So we have an answer to 
the question with which we began. What has happened to the puzzle 
noticed by those nineteenth century giants? It has been transformed 
by some of their twentieth century successors into a puzzle for 
cosmology, a puzzle about the early universe.

It is far from clear how this remaining cosmological puzzle is 
to be explained. Indeed, there are some who
doubt whether it  needs explaining.\cite{call97,call98,sklar93} But these
issues are beyond the scope of this  paper. I want to close by calling
attention to some open questions associated with 
this understanding of  the origins of the thermodynamic asymmetry, and by making a case for an unusually sceptical attitude to the second law.

One fascinating question is whether whatever explains why the 
universe is smooth after the big bang would also imply that the 
universe would be smooth before the big crunch, if the universe 
eventually recollapses. In other words, would entropy would eventually 
decrease, in a recollapsing universe? This possibility was first 
suggested by Thomas Gold some forty years ago.\cite{gold} It has often been 
dismissed  on the grounds that a smooth recollapse would require an
incredibly  unlikely `conspiracy' among the components parts of the
universe,  to ensure that the recollapsing matter did not clump into
black  holes. However, as we have already noted, this incredible
conspiracy  is precisely what happens towards (what we usually term) the
big  bang, if we regard that end of the universe as a product of a 
gravitational collapse. The statistics themselves are time-symmetric. 
If something overrides them at one end of the universe, what 
right do we have to assume that the same does not happen at the 
other? Until we understand more about the origins of the smooth 
early universe, then, it seems best to keep an open mind about a smooth 
late universe.

Another interesting and open question is whether a future 
low entropy boundary condition would have effects \textit{now.} Events 
at the present era provide us with evidence of a low entropy 
past. Could there also be evidence of a low entropy future? The 
answer depends on our temporal distance from such a future 
boundary condition, in relation to the relaxation time of cosmological 
processes. It has been argued that a symmetric time-reversing 
universe would require more radiation in the present era than 
we actually observe---radiation which in the reversed time sense 
originates in the stars and galaxies of the opposite end of the 
universe.\cite{gell} But because of its anti-thermodynamic character,
from  our point of view, it is doubtful whether this radiation would 
be detectable, at least by standard means.\cite{hp1}

Some people dismiss the question whether entropy would reverse 
in a recollapsing universe on the grounds that the current evidence 
suggests that the universe will not recollapse. However, it seems 
reasonable to expect that when we find out why the universe is 
smooth near the big bang, we'll be able to ask a theoretical 
question about what that reason would imply in the case of universe 
which did recollapse. Moreover, as a number of writers have pointed 
out,\cite{hawking,penrose1} much the same question arises if just a bit
of the universe  recollapses---e.g., a galaxy, collapsing into a black
hole. This  process seems to be a miniature version of the gravitational 
collapse of a whole universe, and so it makes sense to ask whether 
whatever constrains the big bang also constrains such partial 
collapses.

\section{Scepticism about the second law}

In my view, the moral of these considerations is that until we 
know more about why entropy is low in the past, it is sensible 
to keep an open mind about whether it might be low in the future. 
The appropriate attitude is a kind of healthy scepticism about 
the universality of the second law of thermodynamics.

The case for scepticism goes like this. What we've learnt about 
why entropy increases in our region is that it does so because 
it is very low in the past (for some reason we don't yet know), 
and the increase we observe is the most likely outcome consistent 
with that restriction. As noted, however, the statistics underpinning 
this reasoning are time-symmetric, and hence the predictions 
we make about the future depend implicitly on the assumption 
that there is no corresponding low entropy boundary condition 
in that direction. Thus the Boltzmann probabilities don't enable 
us to predict without qualification that entropy is unlikely 
to decrease, but only that it is unlikely to decrease, \textit{unless 
there is the kind of boundary condition in the future that makes 
entropy low in the past.} In other words, the second law is likely 
to continue to hold so long as there isn't a low entropy boundary 
condition in the future. But it can't be used to exclude this 
possibility---even probabilistically!

Sceptics about the second law are unusual in the history of 
thermodynamics, and I would like to finish by giving some long-overdue credit to one of the 
rare exceptions. Samuel Hawksley Burbury (1831--1911) was not
one of the true giants of thermodynamics. However, he made  an important contribution to the
identification of the puzzle  of the time-asymmetry of thermodynamic
phenomena. And he was more insightful than any of his
contemporaries---and most writers since, for that matter---in being  commendably cautious about declaring the puzzle
solved.

Burbury  was an English barrister. He read mathematics at Cambridge as an
undergraduate, but his major work in mathematical physics came
late  in life, when deafness curtailed his career at the Bar. In his
sixties and seventies, he thus  played an important in discussions about  the
nature and origins of the second law. 
In a review of Burbury's monograph
\textit{The Kinetic Theory of Gases} for
\textit{Science} in 1899,  the reviewer describes his contribution as follows:

\begin{quote}
[I]n that very interesting discussion of the Kinetic Theory which 
was begun at the Oxford meeting of the British Association in 
1894 and continued for months afterwards in \textit{Nature,} Mr.~Burbury 
took a conspicuous part, appearing as the expounder and defender 
of Boltzmann's H-theorem in answer to the question which so many 
[had] asked in secret, and which Mr.~Culverwell asked in print, `\textit{What 
is the H-theorem and what does it prove?'} Thanks to this discussion, 
and to the more recent publication of Boltzmann's \textit{Vorlesungen 
\"{u}ber Gas-theorie,} and finally to this treatise by Burbury, 
the question is not so difficult to answer as it was a few years 
ago.\cite{hall}

\end{quote}

It is a little misleading to call Burbury a defender 
of the \textit{H}-theorem. The crucial issue in the debate referred to here was the source of the time-asymmetry of the \textit{H}-theorem, 
and while Burbury was the first to put his finger on the role 
of assumption of molecular chaos, he himself regarded this assumption 
with considerable suspicion. Here's how he puts it in 1904:

\begin{quote}
Does not the theory of a general tendency of entropy to diminish 
[\textit{sic}]\footnote{Burbury is apparently
referring to Boltzmann's quantity
\textit{H,}  which does decrease as entropy increases.} take too much for
granted?  To a certain extent it is supported by experimental evidence. 
We must accept such evidence as far as it goes and no further. 
We have no right to supplement it by a large draft of the scientific 
imagination.\cite{burbury04}
\end{quote}

Burbury's reasons for scepticism are not precisely those which 
seem appropriate today. Burbury's concern might be put like this. To 
see that
the dynamical processes routinely fail to produce entropy increases
towards the past is to see that it takes an extra ingredient to ensure
that they do so towards the future. We're then surely right to wonder
whether that extra ingredient is sufficiently universal, even towards the
future, to guarantee that the second law will always hold. As
the first clearly to identify the source of the time-asymmetry in
the
\textit{H}-theorem, Burbury was perhaps more sensitive to this concern
than any of his contemporaries. 

At the same time, however, Burbury seems never to have distanced
himself  sufficiently from the \textit{H}-theorem to see that the real
puzzle of the thermodynamic asymmetry lies  elsewhere. The interesting
question is not whether there is a good dynamical argument to show that
entropy will alway
increase towards the future. It is why entropy steadily
\textit{decreases} towards the past---in the face, note,  of such things
as the effects of collisions and external influences, which
are `happening' in that direction as much as in the other! 
As we've seen, this
re-orientation provides a new reason for being cautious about proclaiming
the universal validity of the second law. Once we regard the fact
that entropy decreases towards the past as itself a puzzle, as something
in need of explanation, then it ought to occur to us that whatever
explains it might be non-unique---and thus that in principle, there
might be a low entropy boundary condition in the future, as well as in
the past.

\end{document}